\documentclass[conference]{IEEEtran}
\IEEEoverridecommandlockouts
\usepackage{amsfonts}
\usepackage{amsmath}
\usepackage{balance}
\usepackage{cite}
\usepackage{hyperref}

\usepackage{ifpdf}
\ifCLASSINFOpdf
   \usepackage[pdftex]{graphicx}
   \graphicspath{{./figures/pdf/}}
   \DeclareGraphicsExtensions{{.pdf}}
\else
   \usepackage[dvips]{graphicx}
   \graphicspath{{./figures/eps/}}
   \DeclareGraphicsExtensions{{.eps}}
\fi

\usepackage[usenames]{color}

\begin{document}
\hyphenation{multi-symbol}
\title{ Multihop Routing in Ad Hoc Networks}
\author{Don Torrieri\IEEEauthorrefmark{1}, \IEEEauthorblockN{ Salvatore Talarico,\IEEEauthorrefmark{2} and Matthew C. Valenti\IEEEauthorrefmark{2}}
\IEEEauthorblockA{\IEEEauthorrefmark{1}U.S. Army Research Laboratory, Adelphi, MD, USA. \\
\IEEEauthorrefmark{2}West Virginia University, Morgantown, WV, USA. }
}
\date{}
\maketitle

\vspace{-1cm}
\thispagestyle{empty}

\title{Multihop Routing in Ad Hoc Networks} \author{\IEEEauthorblockN
{ Don Torrieri,\IEEEauthorrefmark{1} Salvatore Talarico,\IEEEauthorrefmark{2}
and Matthew C. Valenti\IEEEauthorrefmark{2} } \IEEEauthorblockA
{\IEEEauthorrefmark{1}%
U.S. Army Research Laboratory, Adelphi, MD, USA. \\ \IEEEauthorrefmark
{2}West Virginia University, Morgantown, WV, USA.} }%

\begin{abstract}%
This paper presents a dual method of closed-form analysis and lightweight
simulation that enables an evaluation of the performance of
mobile ad hoc networks that is more realistic, efficient, and accurate than
those found in existing publications. Some features
accommodated by the new analysis are shadowing, exclusion and guard zones, and
distance-dependent fading. Three routing protocols are examined: least-delay,
nearest-neighbor, and maximum-progress routing. The tradeoffs among the path
reliabilities, average conditional delays, average conditional number of hops,
and area spectral efficiencies are examined.%
\end{abstract}%

\section{Introduction}

There has been extensive recent research directed toward providing insights
into the tradeoffs among the reliabilities, delays, and throughputs of mobile
ad hoc networks with multihop routing (e.g., \cite{chen} - \cite{and}).
However, the mathematical models and necessary assumptions have not been
adequate for obtaining reliable results. Much of this research uses network
models based on stochastic geometry with the spatial distribution of the
mobiles following a Poisson point process, and simplifying but unrealistic
restrictions and assumptions. One of the principal problems associated with
the models based on stochastic geometry \cite{stoy}, \cite{web} is that they
assume an infinitely large network with an infinite number of mobiles so that
routing in the interior of the network cannot be distinguished from routing
that includes a source or destination mobile near the perimeter of the
network. The Poisson point process does not account for the dependencies in
the placement of mobiles, such as the existence of exclusion zones \cite{tor}
that ensure a minimum spatial separation between mobiles. Among the
unrealistic restrictions are the absence of shadowing, the neglect of thermal
noise, and the identical fading statistics for each link. Among the
unrealistic assumptions are the independence of the success probabilities of
paths from the source to the destination even when paths share the same links,
and the constraining of the number of end-to-end retransmissions rather than link retransmissions.

This analysis in this paper combines
closed-form calculations of outage
probabilities, derived in \cite{tor}, with simple and
rapid simulations that accomodate additional network features.
This approach allows routing to be statistically characterized
without requiring assumptions about
the statistical independence of possible paths. During each
trial of the simulation, the topology is fixed and mobiles are
 placed according to any distribution; we focus on
uniform clustering with exclusion and guard zones.
 Paths for message delivery are selected
by using the closed-form expression for the per-link outage
probability to determine which paths are possible, and the
delay associated with each available link is determined. The number
of transmission attempts per link is constrained, and several routing
protocols are considered. Using these paths and averaging over many
topologies, three routing protocols are examined: least-delay,
nearest-neighbor, and maximum-progress routing. The dependence of the path
reliability, area spectral efficiency, average message delay, and average
number of hops on network parameters such as the maximum number of
transmission attempts per link, source-destination distance, and density of
mobiles are evaluated.

Among the features of our analysis that distinguish it from those by other
authors are the following.

1. Distinct links do not necessarily experience identically distributed fading.

2. Source-destination pairs are not assumed to be stochastically equivalent.
For example if a source or destination is located near the perimeter of the
network, the routing characteristics are different from those computed for
source-destination pairs near the center of the network.

3. There is no assumption of independent path selection, path success
probabilities, or link (hop) success probabilities. Many routes share links,
which cause all these routes to fail if one of the shared links fails. Link
success probabilities are correlated to the degree that lengths of the links
are similar.

4. The shadowing over the link from one mobile to another can be modeled
individually, as required by the local terrain. For computational simplicity
in the examples, the shadowing is assumed to have a lognormal distribution.

5. The analysis accounts for the thermal noise, which is an important consideration
when the mobile density, and
hence the interference, is moderate or low.

6. The routing protocols do not depend on predetermined routes. Instead, they
use the more realistic \emph{dynamic route selection} that entails hop-by-hop
route selection and allows for the possibility of successful communication
over alternative routes.


\section{Network Model}
\label{Section:SystemModel}
The network comprises $M+2$ mobiles in a circular area with radius
$r_{\mathsf{net}}$, although any arbitrary two- or three-dimensional regions
could be considered. The variable $X_{i}$ represents both the $i^{th}$ mobile
and its location, and $||X_{i}-X_{j}||$ is the distance from the $i^{th}$
mobile to the $j^{th}$ mobile. Mobile $X_{0}$ serves as the reference
transmitter or message source, and mobile $X_{M+1}$ serves as the reference
receiver or message destination. The other $M$ mobiles $X_{1},...,X_{M}$ are
potentially relays or sources of interference. Each mobile uses a single
omnidirectional antenna. The radii of the exclusion zones surrounding the
mobiles are equal to $r_{\mathsf{ex}}.$

The source and destination mobiles are placed within the
circular area, and the remaining mobiles $X_1, ..., X_M$ are
uniformly distributed throughout the
network area outside the exclusion zones, according to a \textit{uniform
clustering} model.  One by one, the location of each remaining $X_{i}$ is drawn
according to a uniform distribution within the radius-$r_{\mathsf{net}}$
circle. However, if an $X_{i}$ falls within the exclusion zone of a previously
placed mobile, then it has a new random location assigned to it as many times
as necessary until it falls outside all exclusion zones.
Setting the exclusion zone to $r_{\mathsf{ex}}=0$ is equivalent to drawing the mobiles
from a binomial point process.

In a DS-CDMA network of asynchronous quadriphase direct-sequence systems, a
multiple-access interference signal with power $I$ before despreading is
reduced after despreading to the power level $Ih(\tau_{o})/G$, where G is the
\emph{processing gain} or \emph{spreading factor}, and $h(\tau_{o})$ is a
function of the chip waveform and the timing offset $\tau_{o}$ of the
interference spreading sequence relative to that of the desired or reference
signal. If $\tau_{o}$ is assumed to have a uniform distribution over [0,
$T_{c}],$ then the expected value of $h(\tau_{o})$ is the chip factor $h$. For
rectangular chip waveforms, $h=2/3$ \cite{tor2}, \cite{tor3}. It is assumed
henceforth that $G/h(\tau_{o})$ is a constant equal to $G/h$ at each sector
receiver in the network.

After the despreading, the power of $X_{i}$'s signal at the mobile $X_{j}$ is
\begin{equation}
\rho_{i,j}=\tilde{P}_{i}g_{i,j}10^{\xi_{i,j}/10}f\left(  ||X_{i}%
-X_{j}||\right)  \label{power}%
\end{equation}
where $\tilde{P}_{i}$ is the received power at a reference distance $d_{0}$
(assumed to be sufficiently far that the signals are in the far field) after
despreading when fading and shadowing are absent, $g_{i,j}$ is the power gain
due to fading, $\xi_{i,j}$ is a shadowing factor, and $f(\cdot)$ is a
path-loss function. The path-loss function is expressed as the power law
\begin{equation}
f\left(  d\right)  =\left(  \frac{d}{d_{0}}\right)  ^{-\alpha}\hspace
{-0.45cm},\,\,\text{ \ }d\geq d_{0} \label{pathloss}%
\end{equation}
where $\alpha\geq2$ is the path-loss exponent. It is assumed that
$r_{\mathsf{ex}}\geq d_{0}.$

The \{$g_{i,j}\}$ are independent with unit-mean, but are not necessarily
identically distributed; i.e., the channels from the different $\{X_{i}\}$ to
$X_{j}$ may undergo fading with different distributions. For analytical
tractability and close agreement with measured fading statistics, Nakagami
fading is assumed, and $g_{i,j}=a_{i,j}^{2}$, where $a_{i,j}$ is Nakagami with
parameter $m_{i,j}$. When the channel between $X_{i}$ and $X_{j}$ undergoes
Rayleigh fading, $m_{i,j}=1$ and the corresponding $g_{i,j}$ is exponentially
distributed. In the presence of shadowing with a lognormal distribution, the
$\{\xi_{i,j}\}$ are independent zero-mean Gaussian with variance $\sigma
_{s}^{2}$. For ease of exposition, it is assumed that the shadowing variance
is the same for the entire network, but the results may be easily generalized
to allow for different shadowing variances over parts of the network. In the
absence of shadowing, $\xi_{i,j}=0$. It is assumed that the \{$g_{i,j}\}$
remain fixed for the duration of a time interval but vary independently from
interval to interval (block fading). We define $\mu_{i}$ to be the service
probability that mobile $i$ can serve as a relay along a path from a source to
a destination. With probability $p_{i}$, the $i^{th}$ mobile transmits in the
same time interval as the desired signal. The $\{p_{i}\}$ can be used to model
voice-activity factors, controlled silence, or failed link transmissions and
the resulting retransmission attempts. When the $j^{th}$ mobile is in service as a
potential relay, we set $p_{j}=0.$

The instantaneous signal-to-interference-and-noise ratio (SINR) at the mobile
$X_{j}$ of the signal transmitted by relay $X_{k}$ is given by%
\begin{equation}
\gamma_{j}=\frac{\rho_{k,j}}{\displaystyle{\mathcal{N}}+\sum_{i=1,i\neq k}%
^{M}I_{i}\rho_{i,j}} \label{SINR1}%
\end{equation}
where $\mathcal{N}$ is the noise power, and the indicator $I_{i}$ is a
Bernoulli random variable with probability $P[I_{i}=1]=p_{i}$ and
$P[I_{i}=0]=1-p_{i}$.

Since the despreading does not significantly affect the desired-signal power,
the substitution of (\ref{power}) and (\ref{pathloss}) into (\ref{SINR1})
yields
\begin{equation}
\gamma_{j}=\frac{g_{k,j}\Omega_{k,j}}{\displaystyle\Gamma^{-1}+\sum_{i=1,i\neq
k}^{M}I_{i}g_{i,j}\Omega_{i,j}}%
\end{equation}
where
\begin{equation}
\Omega_{i,j}=%
\begin{cases}
10^{\xi_{k,j}/10}||X_{k}-X_{j}||^{-\alpha} & i=k\\
\displaystyle\frac{h{P}_{i}}{GP_{k}}10^{\xi_{i,j}/10}||X_{i}-X_{j}||^{-\alpha}
& i\neq k
\end{cases}
\end{equation}
is the normalized power of $X_{i}$ at $X_{j}$, ${P}_{i}$ is the received power
from $X_{i}$ at the reference distance $d_{0}$ before despreading when fading
and shadowing are absent, and $\Gamma=d_{0}^{\alpha}P_{k}/\mathcal{N}$ is the
SNR when relay $X_{k}$ is at unit distance from mobile $X_{j}$ and fading and
shadowing are absent.

The \emph{outage probability} quantifies the likelihood that the noise and
interference will be too severe for useful communications. Outage probability
is defined with respect to an SINR threshold $\beta$, which represents the
minimum SINR required for reliable reception. In general, the value of $\beta$
depends on the choice of coding and modulation. An \emph{outage} occurs when
the SINR falls below $\beta$.

In \cite{tor}, closed-form expressions were found for the outage probability
conditioned on the particular network geometry and shadowing factors. Let
$\boldsymbol{\Omega}_j=\{\Omega_{0,j},...,\Omega_{M,j}\}$ represent the set of
normalized powers at $X_{j}$. Conditioning on $\boldsymbol{\Omega}_j$, the
\emph{outage probability} of the link from relay $X_{k}$ to receiver $X_{j}$
is
\begin{equation}
\epsilon_{k,j}=P\left[  \gamma_{j}\leq\beta\big|\boldsymbol{\Omega}_j\right]  .
\end{equation}
The conditioning enables the calculation of the outage probability for any
specific network geometry, which cannot be done by models based on stochastic geometry.

Restricting the Nakagami parameter $m_{k,j}$ of the channel between the relay
$X_{k}$ to receiver $X_{j}$ to be integer-valued, the outage probability
conditioned on $\boldsymbol{\Omega}_j$ is found in \cite{tor} to be
\begin{equation}
\epsilon_{k,j}=1-e^{-\beta_{k,j}z}\sum_{s=0}^{m_{k,j}-1}{\left(  \beta
_{k,j}z\right)  }^{s}\sum_{t=0}^{s}\frac{z^{-t}H_{t,j}}{(s-t)!}%
\end{equation}
where $\beta_{k,j}=\beta m_{k,j}/\Omega_{k,j}$, $z=\Gamma^{-1},$
\vspace{-0.20cm}
\begin{align}
H_{t,j} &  =\mathop{ \sum_{\ell_i \geq 0}}_{\sum_{i=1}%
^{M}\ell_{i}=t}\prod_{i=1,i\neq k}^{M}{\mathsf{G}}_{\ell_i}(i,j)
\label{Hfunc}%
\end{align}
the summation in (\ref{Hfunc}) is over all sets of indices that sum to $t$,
\vspace{-0.20cm}
\begin{equation}
\mathsf{G}_{\ell}(i,j)=%
\begin{cases}
1-p_{i}(1-\Psi_{i,j}^{m_{i,j}}) & \mbox{for $\ell=0$}\\
\frac{p_{i}\Gamma(\ell+m_{i,j})}{\ell!\Gamma(m_{i,j})}\left(  \frac
{\Omega_{i,j}}{m_{i,j}}\right)  ^{\ell}\Psi_{i,j}^{m_{i,j}+\ell} &
\mbox{for $\ell>0$}
\end{cases}
\end{equation}
and
\begin{eqnarray}
\Psi_{i,j}
& = &
\left(  \beta_{k,j}\frac{\Omega_{i,j}}{m_{i,j}}+1\right)
^{-1}\hspace{-0.75cm},\hspace{1cm}\mbox{for $i=\{1,...,M\}$. }
\end{eqnarray}

\section{Routing Models}
\label{Section:RoutingProtocols}
There is no fixed optimal path from source $X_{0}$ to destination $X_{M+1}$ in
a network because every path has a nonzero probability of failure or outage.
Link connectivity and end-to-end connectivity are random variables affected by
the vicissitudes of fading. The optimal path, however defined, can change
within milliseconds. There is one single-hop possible path from $X_{0}$ to
$X_{M+1}$. There are $M$ possible paths with 2 hops, $M\left(  M-1\right)  $
possible paths with no return to $X_{0}$ and 3 hops, and $O\left(
M^{H-1}\right)  $ possible paths with no return to $X_{0}$ and $H$ hops.

Three reactive or on-demand routing protocols are considered:
\emph{least-delay,} \emph{nearest-neighbor, and maximum-progress routing}. All
of these protocols only seek routes when needed. The medium-access-control
protocol is Aloha or CSMA with collision avoidance. To simplify the analysis
when CSMA is used, the CSMA guard zones are assumed to coincide with the
exclusion zones.

In the implementation of least-delay routing, a source $X_{0}$ that seeks to
communicate with a destination $X_{M+1}$ floods the network with request
packets to create it. The transmission of request packets through multiple
routes increases the probability that some request packet successfully reaches
the destination the failures of some paths. The path through intermediate
relays followed by the packet that first reaches the destination determines
the \emph{least-delay path, }which conveys subsequent message packets, and
subsequent receptions of other request packets by the destination are ignored.
The least-delay\emph{\ }path\emph{\ }from $X_{0}$ to $X_{M+1}$ causes the
least interference throughout the network due to multiple transmissions of the
same packet in a multihop path from $X_{0}$ to $X_{M+1}$.

In both nearest-neighbor and maximum-progress routing, the next relay in a
path to the destination is dynamically selected at each hop of each packet and
depends on the local configuration of available relays. A \emph{candidate
link} is one that connects available relays and is not experiencing an outage
due to fading or unfavorable propagation conditions. Nearest-neighbor routing
builds the \emph{nearest-neighbor path} by choosing the closest relay that
lies at the end of a candidate link as the next one in the path from $X_{0}$
to $X_{M+1}$. Maximum-progress routing constructs the \emph{maximum-progress
path} by choosing the next relay on the path as the one that lies at the end
of a candidate link and minimizes the remaining distance to the destination.
In a practical implementation of either nearest-neighbor or maximum-progress
routing, a geographic routing protocol would be used. \emph{Geographic
routing} \cite{cad} selects a path to a destination based on the geographic
positions of the potential relays and makes path decisions at each hop. Each
mobile must know its own location and the locations of other mobiles at the
ends of candidate links. Both the nearest-neighbor and maximum-progress paths
are among the successful paths from $X_{0}$ to $X_{M+1}$ traversed by request
packets during the flooding stage of least-delay routing.

Let $\delta(a,b)$ denote the distance between mobile $a$ and mobile $b$. All
routing methods use a \emph{distance criterion} to exclude a link from mobile
$a$ to mobile $b$ as a link in one of the possible paths from $X_{0}$ to
$X_{M+1}$ if $\delta(b,X_{M+1})>\delta(a,X_{M+1})$. These exclusions, which
will eliminate the majority of links, ensure that each possible path has links
that always reduce the remaining distance to the destination. A mobile may not
be able to serve as a relay in a path from $X_{0}$ to $X_{M+1}$ because it is
already receiving a transmission, is already serving as a relay in another
path, is transmitting, or is otherwise unavailable. All links connected to
mobiles that cannot serve as relays are excluded as links in possible paths
from $X_{0}$ to $X_{M+1}.$ Links that have not been excluded are called
\emph{included} links.

We draw a random realization of the network (topology) using the uniform
clustering distribution of mobiles. Each network topology $t$ is used in
$K_{t}$ simulation trials. The modeling of routing entails the identification
of candidate links. Among the links between possible relays, the distance
criterion is used to exclude various links, and we apply our analysis to
determine the outage probability $\epsilon_{i}$ for each included link $i$. A
Monte Carlo simulation uses the outage probabilities as failure probabilities
to determine which of these links provides a successful transmission after $B$
or fewer transmission attempts. Each included link that passes the latter test
is called a \emph{candidate link }and is assigned a delay determined by the
number of transmission attempts $N_{i}$ required for successful transmission,
where $N_{i}\leq B$.
The \emph{candidate paths} from $X_{0}$ to $X_{M+1}$ are paths that can be
formed by using candidate links.

The \emph{delay of candidate link }$i$ is
$
T_{i}=N_{i}T+(N_{i}-1)T_{e}%
$,
where $T$ is the \emph{delay of a transmission over a link}, and $T_{e}$ is
the \emph{excess delay} caused by a retransmission.
The \emph{delay} $T_{s,t}$
of a path from $X_{0}$ to $X_{M+1}$ for network topology $t$ and simulation
trial $s$ is the sum of the link delays in the path:
\begin{equation}
T_{s,t}=\sum\limits_{i\in\mathcal{L}_{s,t}}[N_{i}T+(N_{i}-1)T_{e}]
\end{equation}
where $\mathcal{L}_{s,t}$ is the set of candidate links constituting the path,
and the $\{T_{s,t}\}$ for topology $t$ are sorted in ascending order of delay;
i.e., $T_{s+1,t} \geq T_{s,t}$.
  If there is a routing failure, then
$1/T_{s,t} = 0$.

\subsection{Least-delay routing}

For least-delay routing, the candidate path with the smallest delay from
$X_{0}$ to $X_{M+1}$ is selected as the \emph{least-delay path }from $X_{0}$
to $X_{M+1}$. This path is determined by using the \emph{Djikstra algorithm}
\cite{bru} with the candidate links and the cost of each link equal to the
delay of the link. If there is no set of candidate links that allow a path
from $X_{0}$ to $X_{M+1},$ then a \emph{routing failure} occurs. If there
are $F_{t}$ routing failures for topology $t$ and $K_{t}$ simulation trials,
then the \emph{probability of end-to-end success} or \emph{path}
\emph{reliability} within topology $t$ is
\begin{equation}
R_{t}=1-\frac{F_{t}}{K_{t}}. \label{R1}%
\end{equation}

Among the $K_{t}-F_{t}$ trials with no routing failure, the \emph{conditional}
\emph{average delay} from $X_{0}$ to $X_{M+1}$ is%
\begin{equation}
D_{t}=\frac{1}{K_{t}-F_{t}}\sum\limits_{s=1}^{K_{t}-F_{t}}T_{s,t}.
\end{equation}
If the least-delay path for trial $s$ has $h_{st}$ links or hops, then among
the $K_{t}-F_{t}$ trials with no routing failure, the \emph{conditional}
\emph{average number of hops} from $X_{0}$ to $X_{M+1}$ is%
\begin{equation}
H_{t}=\frac{1}{K_{t}-F_{t}}\sum\limits_{s=1}^{K_{t}-F_{t}}h_{st}.
\end{equation}

Let $\lambda=(M+1)/\pi r_\mathsf{net}^{2}$ denote the density of transmitters in the
network. We define the \emph{normalized} \emph{area spectral efficiency} for
the $K_{t}$ trials of topology $t$ as
\begin{equation}
\mathcal{A}_{t}=\frac{\lambda}{K_{t}}\sum\limits_{s=1}^{K_{t}}\frac{1}%
{T_{s,t}}%
\end{equation}
where the normalization is with respect to the bit rate or bits per channel
use. The normalized
area spectral efficiency is a measure of the end-to-end throughput in the network.

After computing $R_{t},$ $D_{t},$ $H_{t},$ and $\mathcal{A}_{t}$ for $\Upsilon$ network
topologies, we can average over the topologies to compute the \emph{spatial
averages:}%
\begin{align}
\overline{R}  &  =\frac{1}{\Upsilon}\sum\limits_{t=1}^{\Upsilon}R_{t},\text{
\ }\overline{D}=\frac{1}{\Upsilon}\sum\limits_{t=1}^{\Upsilon}D_{t}\nonumber\\
\text{ \ }\overline{H}  &  =\frac{1}{\Upsilon}\sum\limits_{t=1}^{\Upsilon
}H_{t},\text{ \ }\overline{\mathcal{A}}=\frac{1}{\Upsilon}\sum\limits_{t=1}%
^{\Upsilon}\mathcal{A}_{t}. \label{R7}%
\end{align}

\subsection{Nearest-neighbor routing}

The candidate links are used to determine the nearest-neighbor path from
$X_{0}$ to $X_{M+1}$. Starting with the source $X_{0}$, the candidate link
that spans the shortest Euclidean distance is selected as the first link in
the nearest-neighbor path. The shortest link among the candidate links that
are connected to the relay at the end of the previously selected link is added
successively until the destination $X_{M+1}$ is reached and, hence, the
nearest-neighbor path has been determined. If no nearest-neighbor path from
$X_{0}$ to $X_{M+1}$ can be found, a routing failure is recorded. Equations
$\left(  \ref{R1}\right)  -\left(  \ref{R7}\right)  $ are used to determine
the routing characteristics.

\subsection{Maximum-progress routing}

The candidate links are used to determine the maximum-progress path from
$X_{0}$ to $X_{M+1}$. Starting with the source $X_{0}$, the candidate link
with a terminating relay that minimizes the remaining distance to destination
$X_{M+1}$ is selected as the first link in the maximum-progress path. The link
among the candidate links that minimizes the remaining distance and is
connected to the relay at the end of the previously selected link is added
successively until the destination $X_{M+1}$ is reached and, hence, the
maximum-progress path has been determined. If no maximum-progress path from
$X_{0}$ to $X_{M+1}$ can be found, a routing failure is recorded. Equations
$\left(  \ref{R1}\right)  -\left(  \ref{R7}\right)  $ are used to determine
the routing characteristics.

\subsection{Simulation}

In the dual approach of analysis and simulation, the simulation is simple and
rapid because of the closed-form equation for the outage probability. The
simulation allows the compilation of statistical characteristics of routing
without assumptions about the statistical independence of possible paths. The
simulation can be divided into three levels, each of which corresponds to a
nested \emph{for} loop. The outermost loop (Level 1) is run $\Upsilon$
times, once per topology. The next loop (Level 2) is run $\sqrt{K_{t}}$ times per topology,
and the innermost loop (Level 3) is run $\sqrt{K_{t}}$ times for every iteration of level 2,
so that the total number of trials per topology is $K_t$.

Level 1: Topology. The source mobile is placed at the origin, and the
destination mobile is placed a distance $\delta(X_{0},X_{M+1})$ from it. The
other $M$ mobiles are randomly placed according to the uniform clustering process.

Level 2: Service Model. Each of the $M$ mobiles is marked as available as a
relay with probability $\mu_{i}.$

Level 3: Link-Level Simulation. The outage probability at each potential relay
or destination is computed, where each mobile that is not a potential relay is
a source of interference with probability $p_{i}$. By simulating outages, the
candidate links are determined, and the required number of transmissions is
determined for each of these links.

During each simulation trial, the least-delay, nearest-neighbor, and maximum-progress
routes are identified.

\section{Numerical Results}
\label{Section:Results}
Least-delay routing includes both the nearest-neighbor and maximum-progress
paths among the candidate paths that could be selected by the Djikstra
algorithm. Thus, the least-delay path not only has less or equal delay than
the nearest-neighbor and maximum-progress paths but also more or equal path
reliability and area spectral efficiency.

In the subsequent examples of least-delay routing (LDR), nearest-neighbor
routing (NNR), and maximum-progress routing (MPR), distances and times are
normalized by setting $r_\mathsf{net}=1$ and $T=1$. Other fixed parameter values are
$r_{ex}=0.05,$ $T_{e}=1,$ $\Gamma=0$ dB, $M=200$, $K_{t}=10^{4},$ and
$\Upsilon=1000.$ Each power ${P}_{i}$ is equal. A \emph{distance-dependent
fading} model is assumed, where a signal originating at mobile $X_{i}$ arrives
at mobile $X_{j}$ with a Nakagami fading parameter $m_{i,j}$ that depends on
the distance between the mobiles. We set
\begin{equation}
m_{i,j}=%
\begin{cases}
3 & \mbox{ if }\;||X_{j}-X_{i}||\leq r_{\mathsf{f}}/2\\
2 & \mbox{ if }\;r_{\mathsf{f}}/2<||X_{j}-X_{i}||\leq r_{\mathsf{f}}\\
1 & \mbox{ if }\;||X_{j}-X_{i}||>r_{\mathsf{f}}%
\end{cases}
\end{equation}
where $r_{\mathsf{f}}$ is\ the \emph{line-of-sight radius}. The
distance-dependent-fading model characterizes the typical situation in which
nearby mobiles are in each other's line-of-sight, while mobiles farther away
from each other are not.

\begin{figure}[t]
\centering
\includegraphics[width=9cm]{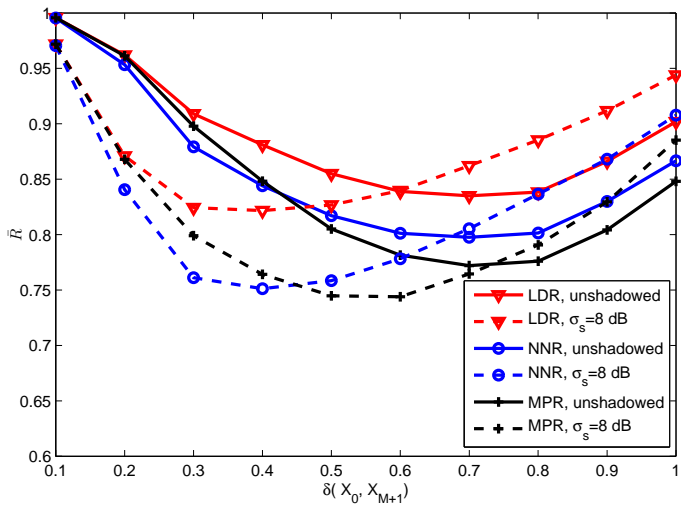}
\vspace{-0.75cm}
\caption{Path reliability as a function of the distance between source and
destination.}
\label{Figure:d_SD_Pe_200}
\vspace{-0.15cm}
\end{figure}

\begin{figure}[t]
\centering
\includegraphics[width=9cm]{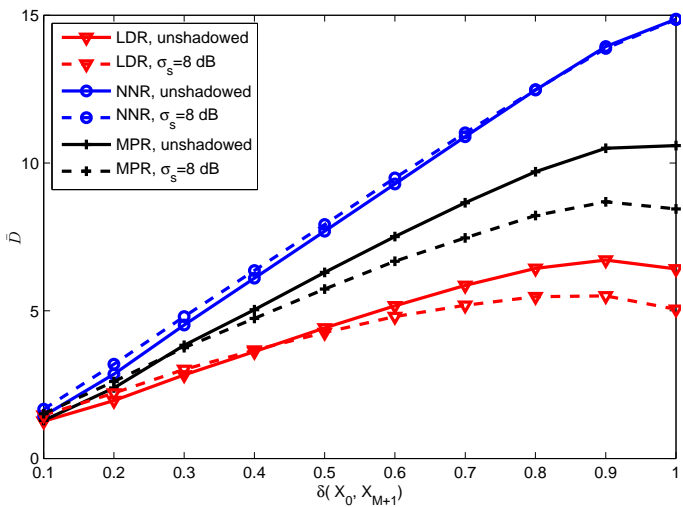}
\vspace{-0.75cm}
\caption{Conditional average delay as a function of the distance between
source and destination.}
\label{Figure:d_SD_D_200}
\vspace{-0.15cm}
\end{figure}

\begin{figure}[t]
\centering
\includegraphics[width=9cm]{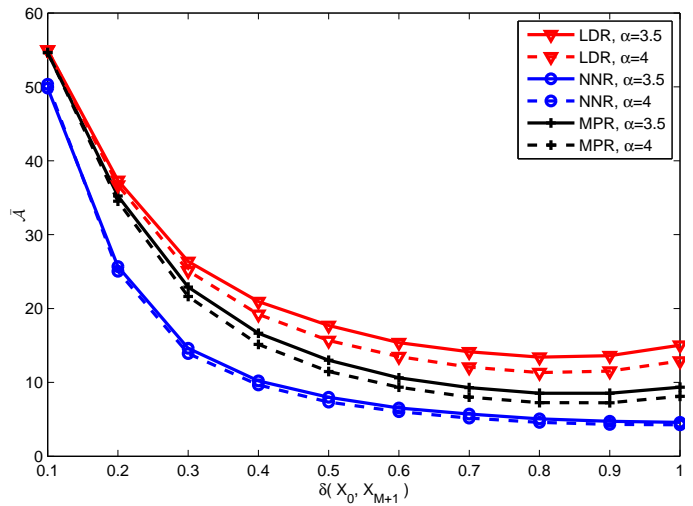}
\vspace{-0.75cm}
\caption{Normalized area spectral efficiency as a function of the distance
between source and destination.}
\label{Figure:d_SD_A_200_alpha}
\vspace{-0.15cm}
\end{figure}

In Figures \ref{Figure:d_SD_Pe_200}, \ref{Figure:d_SD_D_200}, and \ref{Figure:d_SD_A_200_alpha}, $\beta=3$ dB, $B=4$, $G/h=48$, $\mu_{i}=0.3,$
$r_{\mathsf{f}}=0.2,$ and $p_{i}=0.4.$ Figure \ref{Figure:d_SD_Pe_200} shows the variation of path
reliability $\overline{R}$ for each routing algorithm as the distance
$\delta(X_{0},X_{M+1})$ between the source $X_{0}$ and destination $X_{M+1}$
increases. Plots are shown for both no shadowing and lognormal shadowing with
$\sigma_{s}=8$ dB, and $\alpha=3.5$. For smaller values of $\delta
(X_{0},X_{M+1}),$ the shadowing causes a decrease in path reliability, but the
opposite is true for larger values$.$ The increase in path reliability for
larger values of $\delta(X_{0},X_{M+1}),$ particularly with shadowing, is due
to the decreased interference at the destination when it lies on the outskirts
of the network. NNR exhibits greater path reliability than MPR if
$\delta(X_{0},X_{M+1})>0.45.$

\begin{figure}[t]
\centering
\includegraphics[width=9cm]{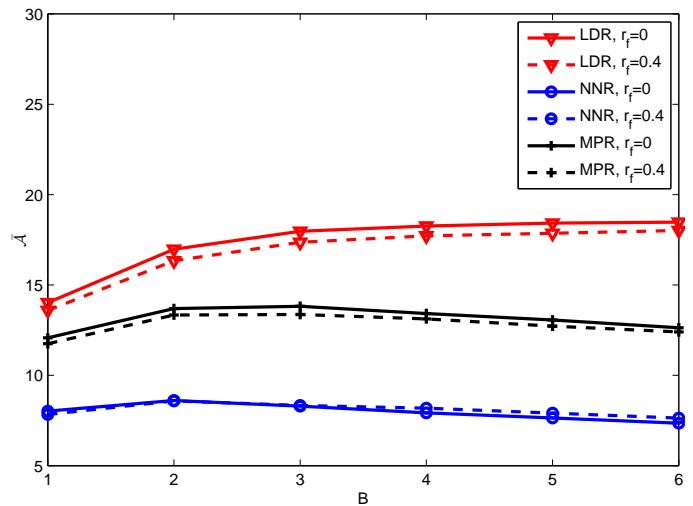}
\vspace{-0.75cm}
\caption{Normalized area spectral efficiency as a function of the number of
allowed transmissions.}
\label{Figure:B_A_200_rf}
\vspace{-0.15cm}
\end{figure}

Figure \ref{Figure:d_SD_D_200} shows the conditional average delay $\overline{D}$ for each routing
algorithm and $\alpha=3.5$ as a function of $\delta(X_{0},X_{M+1}).$ For LDR
and MPR and larger values of $\delta(X_{0},X_{M+1}),$ $\overline{D}$ ceases to
increase, and the shadowing causes an decrease in $\overline{D}.$ For NNR,
$\overline{D}$ increases monotonically with $\delta(X_{0},X_{M+1}),$ and the
shadowing has a relatively small effect. For $\delta(X_{0},X_{M+1})=1$ and
shadowing, it is found that \ $\overline{H}\approx2.0,$ $2.4$, and $7.3$ for
LDR, MPR, and NNR, respectively.

Figure \ref{Figure:d_SD_A_200_alpha} shows the normalized area spectral efficiency $\overline{\mathcal{A}%
}$ for each routing algorithm as a function of $\delta(X_{0},X_{M+1})$ with
$\alpha$ as a parameter. Shadowing with $\sigma_{s}=8$ dB is assumed. As
$\alpha$ increases, both the desired signal and the interference signals are
attenuated. However, particularly for large $\delta(X_{0},X_{M+1}),$ the net
effect is that the more severe propagation conditions reduce $\overline
{\mathcal{A}},$ which is largest for LDR and least for NNR.%

When the number of allowed transmission attempts $B$ increases, the path
reliability and conditional average delay increase for all three routing
algorithms. Figure \ref{Figure:B_A_200_rf} shows the normalized area spectral efficiency
$\overline{\mathcal{A}}$ for each routing algorithm as a function of $B$ with
the line-of-sight radius $r_{\mathsf{f}}$ as a parameter. The parameter values
are $\beta=3$ dB, $\alpha=3.5,$ $\delta(X_{0},X_{M+1})=0.5,$ $G/h=48$,
$\mu_{i}=0.3,$ and $p_{i}=0.4.$ Little or no increase in $\overline
{\mathcal{A}}$ occurs beyond $B=2$. The reason is that more allowed
transmissions not only provide greater path reliability but also lead to
successful paths with longer delays. An increase in $r_{\mathsf{f}}$ is
slightly detrimental.

\begin{figure}[t]
\centering
\includegraphics[width=9cm]{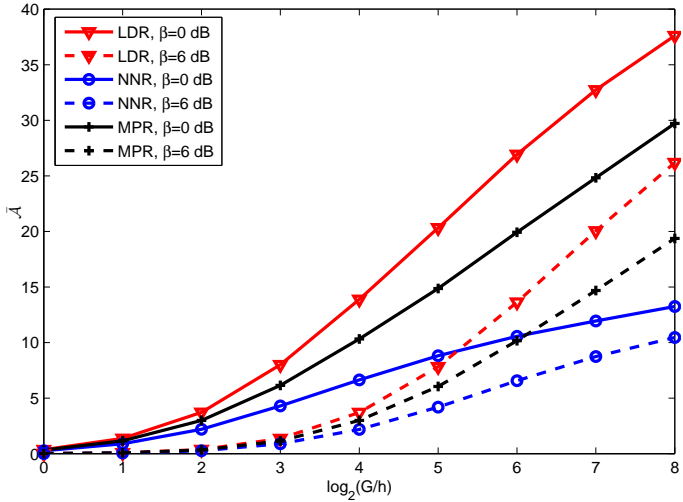}
\vspace{-0.75cm}
\caption{Normalized area spectral efficiency as a function of the spreading
factor.}
\label{Figure:G_A_200}
\vspace{-0.15cm}
\end{figure}

The effects of the spreading factor $G$ and the SINR threshold $\beta$ are
illustrated in Figure \ref{Figure:G_A_200}. All other parameter values are the same as those in
Figure \ref{Figure:B_A_200_rf} except that $r_{\mathsf{f}}=0.2,$ and $B=4.$ The normalized area spectral
efficiency $\overline{\mathcal{A}}$ increases monotonically with G. An
increase in $\beta$ affects LDR\ and MPR more than NNR.

\begin{figure}[t]
\centering
\includegraphics[width=9cm]{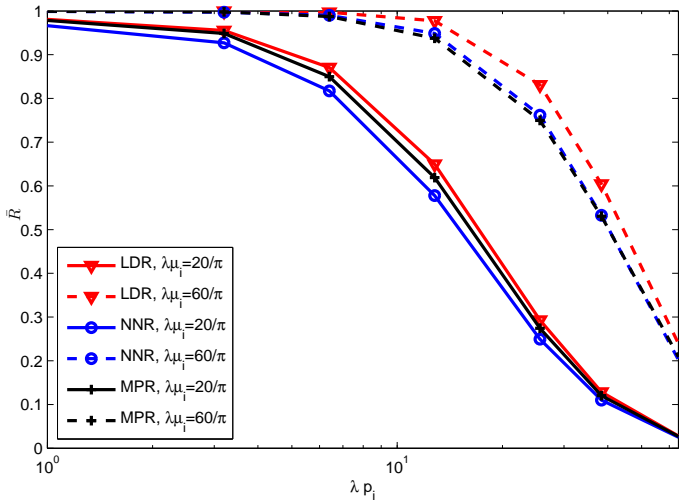}
\vspace{-0.75cm}
\caption{Path reliability as a function of contention density.}
\label{Figure:Pi_Pe_200}
\vspace{-0.15cm}
\end{figure}

Figure \ref{Figure:Pi_Pe_200} shows the path reliability for each routing protocol as a function of
the \emph{contention density\ }$\lambda p_{i}$ with \emph{relay density}
$\lambda\mu_{i}$ as a parameter. This figure indicates the degree to which an
increase in the contention density is mitigated by an increase in relay
density.

\balance

\section{Conclusions}
\label{Section:Conclusion}
This paper presents a new analysis of multihop routing in ad hoc networks.
Many unrealistic and improbable assumptions and restrictions of existing
analyses are discarded. The new analysis is combined with a simulation to
compare three routing protocols: least-delay, nearest-neighbor, and
maximum-progress routing. The tradeoffs among the path reliabilities, average
delays, and area spectral efficiencies of these three routing protocols and
the effects of various parameters have been shown. Least-delay routing has
superior characteristics, but its overhead cost is large because it requires
flooding. Maximum-progress and nearest-neighbor routing both have much lower
overhead costs, but provide more conditional average delay, less area spectral
efficiency, and less path reliability than least-delay routing.
Maximum-progress routing provides less conditional average delay and more area
spectral efficiency than nearest-neighbor routing, but provides less path
reliability when the paths are long.


\begin{thebibliography}{99}                                                                                               %


\bibitem {chen}Y. Chen and J. G. Andrews, "An Upper Bound on Multihop
Transmission Capacity With Dynamic Routing Selection," \textit{IEEE Trans.
Inform. Theory}, vol. 58, pp. 3751-3765, June 2012.

\bibitem {nard2}P. H. J. Nardelli, M. Kaynia, P. Cardieri, and M. Latva-aho,
"Optimal Transmission Capacity of Ad Hoc Networks with Packet
Retransmissions," \textit{IEEE Trans. Wireless Commun.}, vol. 11, pp.
2760-2766, Aug. 2012.

\bibitem {nard}P. H. J. Nardelli, and M. Latva-aho, "Efficiency of Wireless
Networks under Different Hopping Strategies," \textit{IEEE Trans. Wireless
Commun.}, vol. 11, pp. 15-20, Jan. 2012.

\bibitem {vaze}R. Vaze, \textquotedblleft Throughput-delay-reliability
tradeoff with ARQ in wireless ad hoc networks,\textquotedblright\ \textit{IEEE
Trans. Wireless Commun.}, vol. 10, pp. 2142--2149, July 2011.

\bibitem {and}J. G. Andrews, S. Weber, M. Kountouris, and M. Haenggi, "Random
access transport capacity," \textit{IEEE Trans. Wireless Commun.}, vol. 9, pp.
2101-2111, June 2010.

\bibitem {stoy}D. Stoyan, W. Kendall, and J. Mecke, \textit{Stochastic
Geometry and its Applications, 2nd ed.,} Wiley, 1996.

\bibitem {web}S. Weber, J. G. Andrews, and N. Jindal, \textquotedblleft An
overview of the transmission capacity of wireless networks,\textquotedblright%
\ \textit{IEEE Trans. Commun.}, vol. 58, pp. 3593--3604, Dec. 2010.

\bibitem {tor}D. Torrieri and M. C. Valenti, \textquotedblleft The outage
probability of a finite ad hoc network in Nakagami fading,\textquotedblright%
\ \textit{IEEE Trans. Commun.}, vol. 60, pp. 3509-3518, Nov. 2012.

\bibitem {tor2}D. Torrieri, \textit{Principles of Spread-Spectrum
Communication Systems, 2nd ed.} Springer, 2011.

\bibitem {tor3}D.~Torrieri, \textquotedblleft Performance of direct-sequence
systems with long pseudonoise sequences," \emph{IEEE J. Selected Areas
Commun.}, vol.~10, pp.~770-781, May 1992.

\bibitem {cad}F. Cadger, K. Curran, J. Santos, and S. Moffett, "A Survey of
Geographical Routing in Wireless Ad-Hoc Networks," \textit{IEEE Commun.
Surveys Tut.}, vol. 15, pp. 621-653, second quarter, 2013.

\bibitem {bru}R. A. Brualdi, \textit{Introductory Combinatorics, 5th ed.},
Pearson Prentice Hall, 2010.
\end{thebibliography}
\end{document}